# Erratum: Size-dependent piezoelectricity and elasticity in nanostructures due to the flexoelectric effect
# [PHYSICAL REVIEW B 77, 125424 (2008)]

M. S. Majdoub[1], P. Sharma[1,2*], and T. Çağin[3]

[1]Department of Mechanical Engineering, [1]Department of Physics, University of Houston
[3]Department of Chemical Engineering, Texas A&M University

An error was found in the constitutive equations (19) and (20) in our published manuscript: Ref. [1]. Here we correct them. The complete Ref [1] is also included in the discussion of this erratum. The central conclusions of our work (e.g. enhanced size-dependent piezoelectricity in nanostructures due to flexoelectricity) remain the same. In fact, corrected theoretical results in the case of BaTiO$_3$ in piezoelectric phase compare better with atomistics than the original publication [1]. We provide here the corrected equations and for completeness, the revised figures as well.

The correct constitutive equations are:

$$\sigma_{11} = Y\ S_{11} + d\ P_3 - f\ P_{3,3} \tag{1}$$

$$E_3 = a\,P_3 + d\ S_{11} + f\,'S_{11,3} \tag{2}$$

By means of Poisson's equation and in the absence of free charges and applied voltage (open circuit condition) the electric displacement is:

$$D_3 = \varepsilon_0 E_3 + P_3 = 0 \tag{3}$$

Equations (2) and (3) lead to:

$$-(\varepsilon_0^{-1} + a)P_3(x,z) = d\ S_{11} + f\,'S_{11,3} \tag{4}$$

Hence, the correct effective beam bending rigidity (Equation (38) in Ref. [1]) becomes:

$$G = YI[1 + \frac{d^2}{(\varepsilon_0^{-1} + a)Y} + \frac{Aff\,'}{(\varepsilon_0^{-1} + a)YI}] \tag{5}$$

The beam bending rigidity has the elastic, the piezoelectric and the size dependent flexoelectric contributions.

The effective electromechanical coupling factor $k_{eff}$ can be defined from energy consideration as the square root of the ratio of the convertible energy (electric energy) to the total input energy (mechanical energy) (se e.g. References [2,3]).

---

* Corresponding author: psharma@uh.edu

$$k_{eff}^2 = \frac{W_{elec}}{W_{mech}} = \frac{\frac{1}{2}\int \varepsilon E_3^2 dv}{\frac{1}{2}\int Y S_{11}^2 dv} \qquad (6)$$

By means of Equations (2) and (4), the effective electromechanical coupling factor $k_{eff}$ reduces to:

$$k_{eff} = \frac{\chi}{1+\chi}\sqrt{\frac{\varepsilon}{Y}(d^2 + 12(\frac{f'}{h})^2)} \qquad (7)$$

Hence, the normalized effective piezoelectric constant (with bulk piezoelectric constant) is:

$$\frac{d_{eff}}{d_{piez}} = \frac{k_{eff}}{k_{piez}} = \sqrt{(1+12(\frac{f'}{dh})^2)} \qquad (8)$$

We used the flexoelectric constants values estimated by one of us [4] from *ab initio* calculations on BaTiO$_3$ (BT) as $f_{BT} = 5.46\ \dfrac{\text{nC}}{\text{m}}$. The piezoelectric constant of BT is taken from Ref. [5] $d_{BT} = -4.4\ \dfrac{\text{C}}{\text{m}^2}$.

We note that the flexoelectric constants values from *ab initio* calculations are three orders of magnitude lower than the experimental estimates reported by Cross *et al.* [6]. In addition, the existence of such large discrepancy between the *ab initio* calculations [4] and the experimental values [6] was also confirmed by the work of another independent group from Cambridge [7]. The possible reasons behind this discrepancy are discussed in details in Ref. [4].

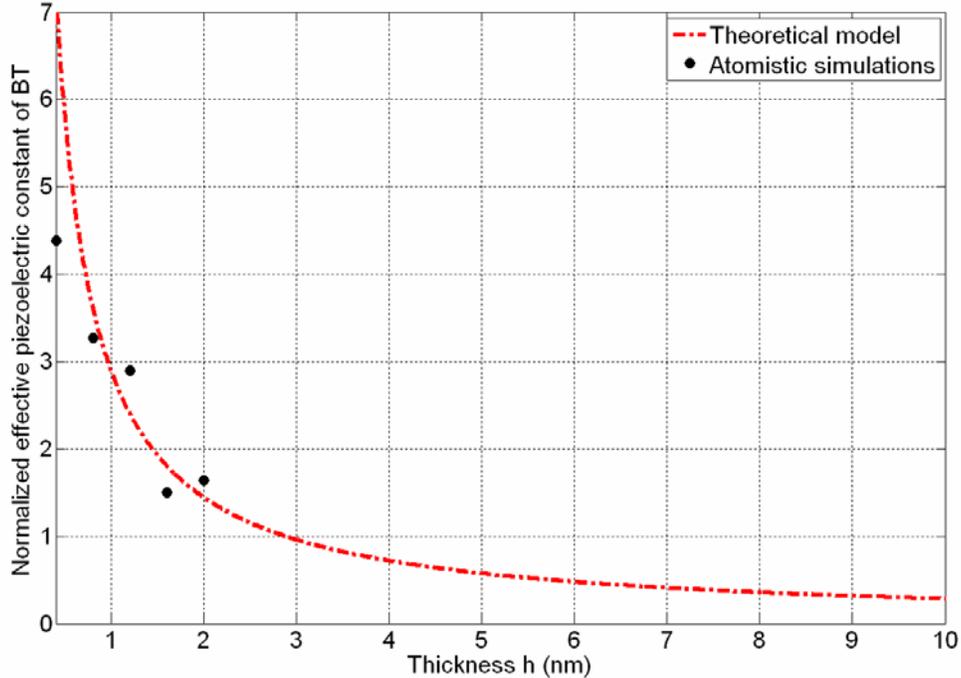

**Figure 1.** Normalized effective piezoelectric constant of cubic (non-piezoelectric) BT. The atomistic simulations are in good agreement with the theoretical model.

The piezoelectric-flexoelectric interaction term incorrectly found in Ref. [1] vanishes in the revised solution. Hence, our non piezoelectric results in Ref. [1] remain valid and the size-dependent behavior due to pure flexoelectricity is seen at the nanoscale (in good agreement with atomistic simulations see Figure 1). However, in the piezoelectric case (see Figure 2), the size dependency is also due to the existence of flexoelectricity and is appreciable down to few nanometers instead of micrometers (as was found in Ref. [1]).

In Figure (2), the effective piezoelectric response is between 3 to 4 times the bulk values at sizes around 2nm. Our theoretical results show that the effective piezoelectric response is doubled at 2nm and increase up to 4 times the piezoelectric constant at 1nm. Thus, the atomistic simulations are able to capture the same order of magnitude as our theoretical results. The corrected results compare better with atomistics (see Figure 9 in [1]). The enhancement in the piezoelectric response is significant but it is only appreciable down to few nanometers.

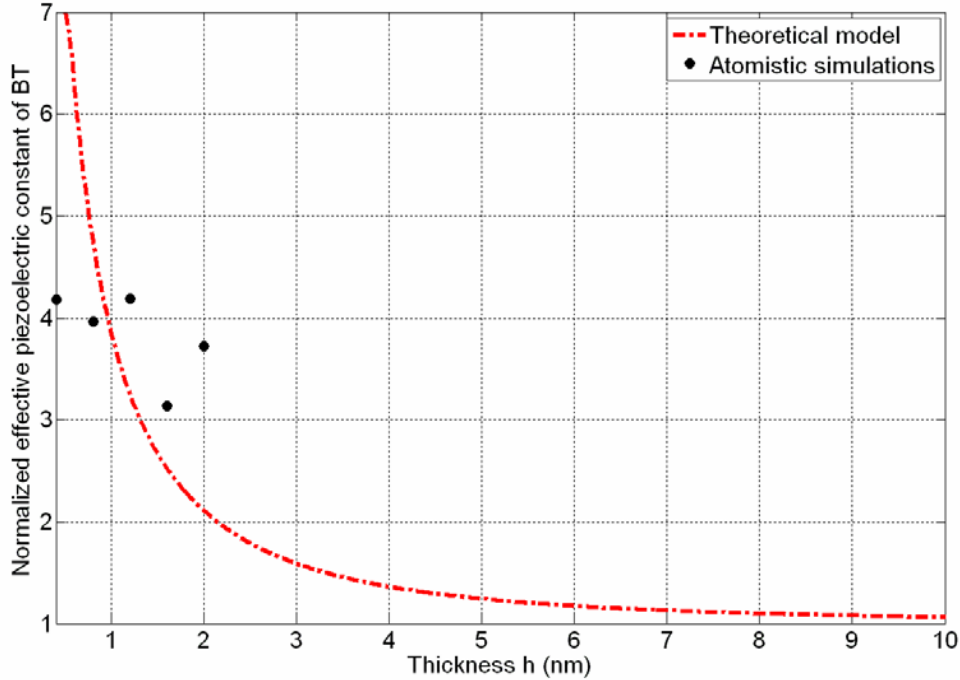

**Figure 2.** Normalized effective piezoelectric constant of tetragonal (piezoelectric) BT. The atomistic simulations are able to capture the same order of magnitude as the theoretical model. The enhancement in the piezoelectric response is seen at the nanosize.

The results will change if flexoelectric values of [6] are used instead of first principle calculations. It is a puzzle that various atomistic models agree with each other (e.g. 4, 7) and so do experiments (e.g. 6, 7) but atomistic models do not agree with experiments for ferroelectrics. This issue remains unresolved and meanwhile to be consistent, when comparing our theoretical results to atomistics we used the flexoelectric properties from atomistics (and likewise, comparison of various experimental results should be predicated on experimental flexoelectric values).

*The authors are very grateful to A. K. Tagantsev. G. Catalan, R. Maranganti and M. Gharbi for valuable discussions.*

# Size-dependent super-piezoelectricity and elasticity in nanostructures due to the flexoelectric effect

M. S. Majdoub[1], P. Sharma[1,2,*] and T. Cagin[3]

[1]Department of Mechanical Engineering, University of Houston, Houston, TX, 77204, U.S.A
[2]Department of Physics, University of Houston, Houston, TX, 77204, U.S.A
[3]Department of Chemical Engineering, Texas A&M University, College Station, TX 77845, U.S.A

**Abstract:** Crystalline piezoelectric dielectrics electrically polarize upon application of uniform mechanical strain. Inhomogeneous strain, however, locally breaks inversion symmetry and can potentially polarize even non-piezoelectric (centrosymmetric) dielectrics. Flexoelectricty--the coupling of strain gradient to polarization-- is expected to show a strong size-dependency due to the scaling of stain gradients with structural feature size. In this study, using a combination of atomistic and theoretical approaches, we investigate the "effective" size-dependent piezoelectric and elastic behavior of inhomogeneously strained non-piezoelectric and piezoelectric nanostructures. In particular, to obtain analytical results and tease out the novel physical insights, we analyze a paradigmatic nanoscale cantilever beam. We find that in materials that are intrinsically piezoelectric, the flexoelectricity and piezoelectricity effects do not add linearly and exhibit a nonlinear interaction. The latter leads to a strong size-dependent enhancement of the apparent piezoelectric coefficient resulting in, for example, a "giant" 500% enhancement over bulk properties in BaTiO$_3$ for a beam thickness of 5 nm. Correspondingly, for non-piezoelectric materials also, the enhancement is non-trivial (e.g. 80 % for 5 nm size in paraelectric BaTiO$_3$ phase). Flexoelectricity also modifies the apparent elastic modulus of nanostructures, exhibiting an asymptotic scaling of $1/h^2$ where "h" is the characteristic feature size. Our major predictions are verified by quantum mechanically derived force-field based molecular dynamics for two phases (cubic and tetragonal) of BaTiO$_3$.

## I. INTRODUCTION

In response to mechanical stimuli, certain crystalline dielectrics may electrically polarize. Assuming that the applied **uniform** mechanical strain, $\varepsilon$, is "small enough"[1], empirical evidence and phenomenological considerations suggest the following relation:

$$(\mathbf{P})_i = (\mathbf{d})_{ijk}(\boldsymbol{\varepsilon})_{jk} \tag{1}$$

Indices (in some suitable Cartesian framework) are explicitly written to display the order of the matter tensors as prevalently understood in the literature. The third order tensor **d** is thus the piezoelectric matter tensor. Symmetry considerations restrict it be non-zero only for dielectrics belonging to crystallographic point groups that admit non-centrosymmetry[3].

Centrosymmetric dielectrics evidently are not expected to polarize under mechanical strain. A *non-uniform strain* field or the presence of *strain gradients* can however, locally break inversion symmetry and induce polarization even in centrosymmetric crystals. This phenomenon is termed flexoelectrictiy,[4,5] inspired by a similar effect in liquid crystals[6-8]. In a naïve approach, we may simply append a term proportional to the strain gradients to Equation (1):

$$(\mathbf{P})_i = (\mathbf{d})_{ijk}(\varepsilon)_{jk} + (\mathbf{f})_{ijkl}\nabla_l(\varepsilon)_{jk} \qquad (2)$$

Here $\mathbf{f}$ is the so-called fourth order flexoelectric tensor. Thus, unlike the components of the third ordered tensor '$\mathbf{d}$' (piezoelectric coefficients) which are non-zero for only selected (piezoelectric) dielectrics, the flexoelectric coefficients (components of the fourth order tensor '$\mathbf{f}$') are, *in principle*, non-zero for *all* dielectrics although of course they may be negligibly small for many materials. The reader is referred to Tagantsev[9,10] who provides an overview of the subject. In a recent work, one of us[11] has discussed a mathematical framework for flexoelectricity in detail, in addition to providing a review of this subject.

Recently, flexoelectricity has caught the attention of several researchers and indeed some have proposed tantalizing notions related to this phenomenon. For instance, Cross and co-workers[12] were the first to suggest that flexoelectricity should allow fabrication of "piezoelectric composites without using piezoelectric materials". One of us has computationally analyzed such meta-materials while Cross[13-15] *et al* have fabricated non-piezoelectric tapered pyramidal structures on a substrate that "effectively" act as piezoelectric meta-materials. Flexoelectricity is also seen to play an important role in the characteristics of ferroelectrics e.g. Catalan *et al*[5] study the effect of flexoelectricity on the dielectric constant, polarization and Curie temperature in ferroelectric thin films under in-plane substrate induced epitaxial strain.

Patently, the strength of the flexoelectric size-effects crucially depends upon either the numerical values of the flexoelectric coefficients or how large the strain gradients are. The latter is closely linked with the size-scale of the structure. Consider two embedded triangular inclusions[16] (Figure 1) subject to a stress at two different length scales but with the *same aspect ratios*. While the strain field remains the same across both length scales, the strain gradients scale as $1/a_i$ (where $a_i$ designates a distance between two points inside the inclusion). This simple notion is the essence of the size-effect displayed by flexoelectricity.

Flexoelectric coefficients are not readily available but some reasonable estimates are known for some specific materials e.g. atomistic calculations for graphene (Dumitrica et. al.[18] and Kalinin et. al.[19]) and lattice dynamics for NaCl (Askar and Lee[20]). Kogan[21] has argued that for all dielectrics, $e/a$ ($\approx 10^{-9}$ C/m) is an appropriate lower bound for the flexoelectric coefficients, where $e$ is the electronic charge and $a$ is the lattice parameter. Later experiments (Ma and Cross[22]) and simple linear chain models of ions (Marvan and Havranek[23]) suggested multiplication by relative permittivity for normal dielectrics. Much larger magnitudes ($\approx 10^{-6}$ C/m) of flexoelectric coefficients than this lower bound are observed in certain ceramics[24-26]. Flexoelectricity, of course also exists in dielectrics that are already piezoelectric and in fact experimental evidence suggests that flexoelectric coefficients are unusually high in such materials---see the work of Cross and co-workers[13, 22, 24-26] on ferroelectric perovskites like BST ( $f_{BST} = 100 \mu C/m$ ), PZT ( $f_{PZT} = 0.5 - 2 \mu C/m$ at lower and higher strain gradients), and PMN ( $f_{PMN} = 4 \mu C/m$ ). Here we note that the flexoelectric coefficient of ferroelectric materials is quite high even in the paraelectric phase. Quite remarkably, Zubko *et al*[27] have recently published the experimental characterization of the complete flexoelectric tensor for SrTiO₃.

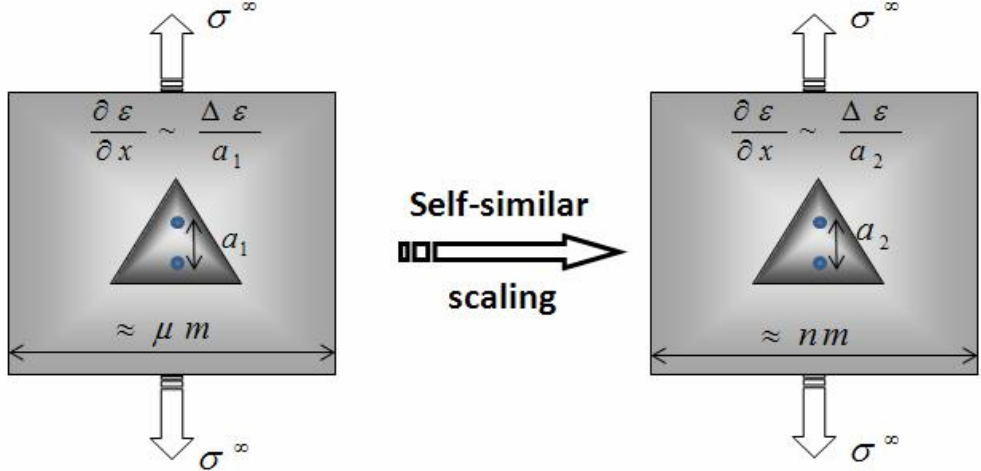

**Figure 1:** Illustration of size-effects due to scaling of strain gradients. Subjected to the same far field stress, two triangular inclusions kept at the *same aspect ratio* but at different length scales will exhibit strain gradients that scale as $1/a_i$.

In the present work, we analyze the role of flexoelectricity in both piezoelectric and non-piezoelectric nanostructures. In particular, we focus on the illustrative model problem of a nanoscale cantilever beam to obtain analytical expressions for the "effective" or "apparent" size-dependent piezoelectric coefficient and elastic modulus. The simplicity of the chosen model system allows a facile inference of various physical insights. On this note, we also observe that cantilever beams have important technological ramifications as actuators, sensors, energy harvesting among others[28-33]. Zhong *et al.*[34] used atomic force microscopy to deflect the tips of aligned arrays of piezoelectric cantilever zinc oxide nanowires. Due to bending, such nano-harvesting devices show generated piezoelectric power efficiency up to 30%. To verify our predictions, we carry out atomistic calculations on both paraelectric and piezoelectric phases of BaTiO₃ (BT) nano cantilever beams under bending deformation.

The paper is organized as follows. In Section II, we summarize the mathematical framework and the governing equations of flexoelectricity. In Section III, we develop solutions for the model nano-scale cantilever beam. Based on the analytical results for this paradigmattical problem, we present the key physical insights in Section IV and in particular discuss the possibility of giant piezoelectricity at the nanoscale and the size-dependent re-normalization of the elastic modulus. In Section V, we present our atomistic calculations and conclude in Section VI.

## II. THEORY OF FLEXOELECTRICITY AND GOVERNING EQUATIONS

In this section, our presentation closely follows the following references[35-37] including one of our recent papers[11]. We note that the correct incorporation of flexoelectricity naturally necessitates the inclusion of polarization gradients also (the latter was first introduced by Mindlin[35]). The symbol *Lin* designates the set of all linear transformations and the associated inner product is defined as: $\langle \mathbf{A}, \mathbf{B} \rangle = tr(\mathbf{A}^T \mathbf{B})$.

For a dielectric occupying a volume V bounded by a surface S in a vacuum $V^{'}$, with a total volume $V^{*}$, Hamilton's principle may be written as:

$$\delta \int_{t_1}^{t_2} dt \int_{V^*} (\frac{1}{2} \rho \langle \dot{\mathbf{u}}, \dot{\mathbf{u}} \rangle - H) dV + \int_{t_1}^{t_2} dt [\int_V (\langle \mathbf{f}, \delta\mathbf{u} \rangle + \langle \mathbf{E}^0, \delta\mathbf{P} \rangle) \delta V + \int_S \langle \mathbf{t}, \delta\mathbf{u} \rangle dS] = 0 \qquad (3)$$

where $\mathbf{u}$, $\mathbf{P}$, $\mathbf{f}$, $\mathbf{E}^0$ and t are respectively the displacement, the polarization, the external body force, electric field and surface traction.

The electric enthalpy density $H$ was defined by Toupin[38] and divided into energy density of deformation and polarization denoted $w^L$ and a reminder. By extending the dependence of $w^L$ to include both strain and polarization gradients, the electric enthalpy density $H$ takes the following form:

$$H = W^L(\mathbf{S}, \mathbf{P}, \nabla\nabla\mathbf{u}, \nabla\mathbf{P}) - \frac{1}{2} \varepsilon_0 \langle \nabla\varphi, \nabla\varphi \rangle + \langle \nabla\varphi, \mathbf{P} \rangle \qquad (4)$$

where $\mathbf{S}$ is the symmetric strain tensor, $\varphi$ is the potential of the Maxwell self-field defined by $\mathbf{E}^{MS} = -\nabla\varphi$ and $\varepsilon_0$ is the permittivity of the vacuum.

Assuming an independent variation of the displacement, polarization, electric potential and their gradients, the variation of the electric enthalpy density $\delta H$ is:

$$\delta H = \langle \mathbf{T}, \delta\mathbf{S} \rangle - \langle \bar{\mathbf{E}}, \delta\mathbf{P} \rangle + \langle \tilde{\mathbf{T}}, \delta\nabla\nabla\mathbf{u} \rangle + \langle \tilde{\mathbf{E}}, \delta\nabla\mathbf{P} \rangle - \varepsilon_0 \langle \nabla\varphi, \delta\nabla\varphi \rangle$$
$$+ \langle \nabla\varphi, \delta\mathbf{P} \rangle + \langle \mathbf{P}, \delta\nabla\varphi \rangle \qquad (5)$$

where,

$$\mathbf{T} = \frac{\partial W^L}{\partial \mathbf{S}}, \quad \bar{\mathbf{E}} = -\frac{\partial W^L}{\partial \mathbf{P}}, \quad \tilde{\mathbf{T}} = \frac{\partial W^L}{\partial \nabla\nabla\mathbf{u}}, \quad \tilde{\mathbf{E}} = \frac{\partial W^L}{\partial \nabla\mathbf{P}} \qquad (6)$$

$\mathbf{T}$ is the stress tensor, $\bar{\mathbf{E}}$ is the effective local electric force, $\tilde{\mathbf{T}}$ and $\tilde{\mathbf{E}}$ can be interpreted as higher order stress and local electric force respectively.

Using the chain rule of differentiation,

$$\delta H = \nabla.(\mathbf{T}.\delta\mathbf{u}) - \langle \nabla.\mathbf{T}, \delta\mathbf{u} \rangle + \nabla.(\tilde{\mathbf{T}}.\delta\nabla\mathbf{u}) - \langle \nabla.\tilde{\mathbf{T}}, \delta\nabla\mathbf{u} \rangle - \langle \bar{\mathbf{E}} - \nabla\varphi, \delta\mathbf{P} \rangle$$
$$+ \nabla.(\tilde{\mathbf{E}}\delta\mathbf{P}) - \langle \nabla.\tilde{\mathbf{E}}, \delta\mathbf{P} \rangle + \nabla.[(-\varepsilon_0\nabla\varphi + \mathbf{P})\delta\varphi] - (-\varepsilon_0\Delta\varphi + \nabla.\mathbf{P})\delta\varphi \qquad (7)$$

The kinetic energy in Equation (3) is written as:

$$\delta \int_{t_1}^{t_2} dt \int_{V^*} \frac{1}{2} \rho \langle \dot{\mathbf{u}}, \dot{\mathbf{u}} \rangle dV = -\int_{t_1}^{t_2} dt \int_{V^*} \rho \langle \ddot{\mathbf{u}}, \delta\mathbf{u} \rangle dV \qquad (8)$$

Substituting Equation (7) into the Hamilton principle Equation (3) and by use of divergence theorem, we find that,

$$\int_{t_1}^{t_2} dt \int_{V^*} [\langle (-\rho\ddot{\mathbf{u}} + \nabla.\mathbf{T} - \nabla.(\nabla.\tilde{\mathbf{T}}) + \mathbf{f}), \delta\mathbf{u} \rangle + \langle (\bar{\mathbf{E}} - \nabla\varphi + \nabla.\tilde{\mathbf{E}} + \mathbf{E}^0), \delta\mathbf{P} \rangle$$

$$+ (-\varepsilon_0\Delta\varphi + \nabla.\mathbf{P})\delta\varphi] dV + \int_{t_1}^{t_2} dt \int_{S} [\langle [(-\mathbf{T} + \nabla.\tilde{\mathbf{T}}).\mathbf{n} + \mathbf{t}], \delta\mathbf{u} \rangle - \langle \tilde{\mathbf{E}}.\mathbf{n}, \delta\mathbf{P} \rangle \quad (9)$$

$$- \langle -\varepsilon_0\nabla\varphi + \mathbf{P}, \mathbf{n} \rangle \delta\varphi] dS = 0$$

Hence, the equilibrium equations are

$$\nabla.\boldsymbol{\sigma} + \mathbf{f} = \rho\ddot{\mathbf{u}} \text{ where } \boldsymbol{\sigma} = \mathbf{T} - \nabla.\tilde{\mathbf{T}} \text{ in V}$$

$$\bar{\mathbf{E}} + \nabla.\tilde{\mathbf{E}} - \nabla\varphi + \mathbf{E}^0 = 0 \text{ in V} \quad (10)$$

$$-\varepsilon_0\Delta\varphi + \nabla.\mathbf{P} = 0 \text{ in V and } \Delta\varphi = 0 \text{ in V}'$$

whereas the corresponding boundary conditions on S are

$$\boldsymbol{\sigma}.\mathbf{n} = \mathbf{t} \text{ where } \boldsymbol{\sigma} = \mathbf{T} - \nabla.\tilde{\mathbf{T}}$$

$$\tilde{\mathbf{E}}.\mathbf{n} = 0 \quad (11)$$

$$(-\varepsilon_0\|\nabla\varphi\| + \mathbf{P}).\mathbf{n} = 0$$

$\boldsymbol{\sigma}$ may be considered as the actual physical stress experienced by a material point and differs from the Cauchy stress $\mathbf{T}$. The symbol $\|\ \|$ denotes the jump across the surface or an interface.

Neglecting the contribution of higher order terms (fifth order tensors and higher)- the strain energy density can be expanded as

$$W^L(\mathbf{S}, \mathbf{P}, \nabla\nabla\mathbf{u}, \nabla\mathbf{P}) = \frac{1}{2}\mathbf{P}.\mathbf{a}.\mathbf{P} + \frac{1}{2}\nabla\mathbf{P}:\mathbf{b}:\nabla\mathbf{P} + \frac{1}{2}\mathbf{S}:\mathbf{c}:\mathbf{S}$$

$$+ \mathbf{S}:\mathbf{e}:\nabla\mathbf{P} + \mathbf{S}:\mathbf{d}.\mathbf{P} + \mathbf{P}.\mathbf{g}:\nabla\mathbf{P} + \mathbf{P}.\mathbf{f}:\nabla\nabla\mathbf{u} \quad (12)$$

Finally, according to Equation (6), the constitutive equations are

$$\mathbf{T} = \frac{\partial W^L}{\partial \mathbf{S}} = \mathbf{c}:\mathbf{S} + \mathbf{e}:\nabla\mathbf{P} + \mathbf{d}.\mathbf{P}$$

$$\tilde{\mathbf{T}} = \frac{\partial W^L}{\partial \nabla\nabla\mathbf{u}} = \mathbf{f}.\mathbf{P}$$

$$-\bar{\mathbf{E}} = \frac{\partial W^L}{\partial \mathbf{P}} = \mathbf{a}.\mathbf{P} + \mathbf{g}:\nabla\mathbf{P} + \mathbf{f}:\nabla\nabla\mathbf{u} + \mathbf{d}:\mathbf{S} \quad (13)$$

$$\tilde{\mathbf{E}} = \frac{\partial W^L}{\partial \nabla\mathbf{P}} = \mathbf{b}:\nabla\mathbf{P} + \mathbf{e}:\mathbf{S} + \mathbf{g}.\mathbf{P}$$

The coefficients of the displacement, polarization and their gradients defined above as "a", "b", "c", "d", "f", "g" and "e" are material property tensors. The second order tensor "a" is the reciprocal dielectric susceptibility. The fourth order tensor "b" is the polarization gradient-polarization gradient coupling tensor and "**c**" is the elastic tensor. The fourth order tensor "**e**" correspond to polarization gradient and strain coupling introduced by Mindlin[33] whereas "**f**" is the fourth order flexoelectric tensor. "**d**" and "**g**" are the third order piezoelectric tensor and the polarization-polarization gradient coupling tensor.

## III. MODEL PROBLEM: CANTILEVER NANO-BEAM

Piezoelectric materials generally have symmetry lower than cubic and (even for the latter) analytical calculations are all but impossible for general three-dimensional bodies. A cantilever beam is a model system that degenerates to a one dimensional problem and is thus analytically tractable (albeit approximately). Figure (2) depicts the schematic of such a cantilever beam. We note that a closed-form solution of a cantilever predicated on classical piezoelectric theory (*excluding* the flexoelectric effect) has been derived by Weinberg[39]. The latter work ignores variation of electric field through the thickness of the beam and accordingly is only valid for materials with low electro-mechanical coupling. Subsequently Tadmor *et al.*[40] have improved upon on that work by taking into account the variation of the electric field in the beam layers.

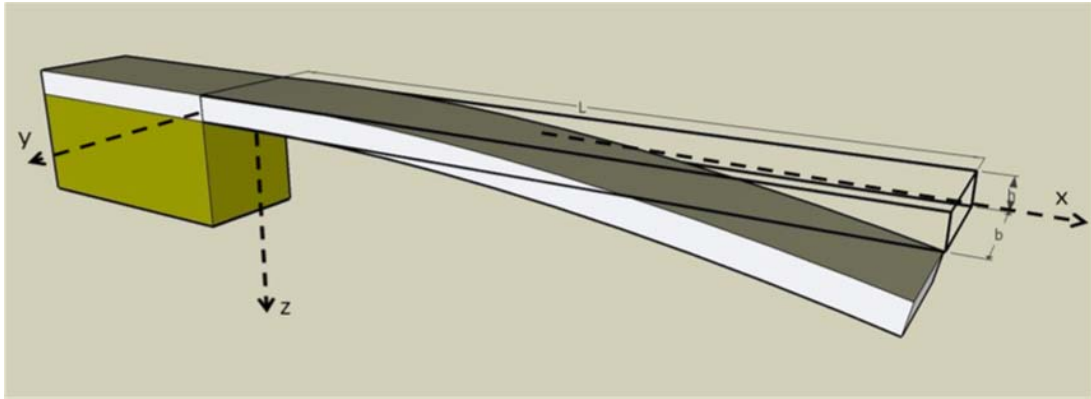

**Figure 2:** Schematic of a rectangular cantilever beam. Initial and bent configurations are sketched.

We adopt the usual assumptions made in analyzing slender beams e.g. beam thickness is much less than the radius of curvature induced by the mechanical and electrical loading and that beam cross section is constant along its length.
In the adopted Oxyz Cartesian coordinate system (Figure 2), Ox corresponds to the centroidal axis of the undeformed beam, y-axis is the neutral axis and the z the symmetry axis. Although a rectangular cross-sectional beam is depicted in Figure (2), much of the derivation proceeds for an arbitrary cross-sectional shape.

The displacement field is, $\mathbf{u} = \mathbf{u}(u_1(x,z), u_2 = 0, u_3(x))$. As typical in the analysis of beams, the displacement is parameterized with respect to the out-of-plane displacement component:

$$u_3 = w(x)$$

$$u_1 = -z\frac{du_3(x)}{dx} = -z\frac{dw(x)}{dx} \tag{14}$$

$$u_2 = 0$$

For narrow beams (b<5h), it is typical to assume that the stresses $\sigma_{33} = \sigma_3 = 0$ and $\sigma_{22} = \sigma_2 = 0$. The only relevant electric field component is $E_3$. According to the physical stress defined in Equation (10), the non-vanishing component $\sigma_{11}$ is:

$$\sigma_{11} = \sigma_1 = T_{11} - T_{111,1} - T_{113,3} \tag{15}$$

Without loss of generality we now assume tetragonal 4mm material symmetry. Most piezoelectrics are of the latter or higher symmetry (e.g. PZT 5H). Accordingly, we can re-write Equation (15) as:

$$\sigma_1 = c_1 S_1 + (e_{13} - f_{13})P_{3,3} + d_{31}P_3 \tag{16}$$

in which the Voigt notation is used for the different coefficients and $c_1$, $d_{31}$, $e_{13}$ and $f_{13}$ designate respectively the elastic modulus, the piezoelectric constant, the polarization gradient and strain coupling constant and the flexoelectric coefficient of the one-dimensional beam.

$S_1$ is the axial strain which can be explicitly written under the beam assumptions as function of the radius curvature R(x):

$$S_1(x,z) = -\frac{z}{R(x)} = -z\frac{d^2w(x)}{dx^2} \tag{17}$$

Equation (14) may be rewritten with a somewhat simpler notation as:

$$\sigma_1 = YS_1 + (e - f)P_{3,3} - Yd\ P_3 \tag{18}$$

Here $Y = c_1$, $e = e_{13}$, $f = f_{13}$ and $d = -d_{31}/Y$ .

The notation in Equation (18) facilitates subsequent comparison with results obtained by Tadmor *et al*[40] for classical piezoelectric beams.

Finally, the electric field induced by the polarization due to piezoelectricity and flexoelectricity (strain gradient term) is expressed as:

$$E_3 = \varepsilon_0^{-1}\chi_{33}^{-1}P_3 - f_{55}S_{11,3} \tag{19}$$

where $\chi_{33}^{-1} = \varepsilon_0 a_{33} = \chi^{-1}$ is the reciprocal dielectric susceptibility.

The total electric displacement in z-direction is given by

$$D_3 = d\sigma_1 + \varepsilon\ E_3 + f_{55}S_{11,3} \tag{20}$$

with $\varepsilon = \varepsilon_{33}$ is the dielectric constant. Predicated on 1-D beam assumptions we have from Equation (20)

$$E(x,z) = \frac{1}{\varepsilon}(D_3(x) - d\sigma_1(x,z) - f_{55}S_{11,3}) \tag{21}$$

We define the "through-layer" average of any quantity as:

$$\overline{T}(x) = \frac{1}{h}\int_{Layer} T(x,z)dz \tag{22}$$

Since the applied voltage difference is constant along the beam,

$$V = -\frac{D_3(x)h}{\varepsilon} + \frac{dh}{\varepsilon}\overline{\sigma}_1 + \frac{f_{55}h}{\varepsilon}\overline{S}_{11,3} \tag{23}$$

We may thus write:

$$D_3(x) = -\frac{\varepsilon V}{h} + d\overline{\sigma}_1 + f_{55}\overline{S}_{11,3} \tag{24}$$

From Equation (18), the average layer stress is then:

$$\overline{\sigma}_1 = (e-f)\overline{P}_{3,3} - Yd\,\overline{P}_3 \tag{25}$$

Assuming a linear variation of the electric field in z and that the average layer electric field and voltage are respectively equal to $-\frac{V}{h}$ and $V$, we find that

$\overline{E}_3 = -\frac{V}{h}$, $\overline{E}_{3,3} = -\frac{24V}{h^2}$ and hence $\overline{P}_3$ and $\overline{P}_{3,3}$.

Substituting Equation (24) into (21) with (25) we obtain an equation to solve for the electric polarization

$$-\frac{V\varepsilon_0\chi}{h} - \frac{d}{\varepsilon}(\sigma_1(x,z) - \overline{\sigma}_1) = P(x,z) - f'\,S_{11,3} \tag{26}$$

in which $f' = f_{55}$.

Solving Equation (26), we obtain:

$$P(x,z) = \underbrace{-\frac{V\varepsilon_0\chi}{h}}_{electrostatic} + \underbrace{\frac{\xi}{d}\frac{z}{R(x)}}_{pure\ piezoelectricity} - \underbrace{\frac{f'}{R(x)}}_{pure\ flexoelectricity} \underbrace{-\frac{\xi^2(e-f)}{d^2Y\,R(x)} - \frac{24V\varepsilon_0\chi\xi(e-f)}{dYh^2}}_{piezoelectricity\text{-}flexoelectricity\ interaction} \tag{27}$$

where $\xi = k_e^2 = \frac{k^2}{1-k^2}$ is defined as the the square of the expedient coupling coefficient[41] $k_e$ and $k = \sqrt{\frac{Yd^2}{\varepsilon}}$ is the so-called Electro Mechanical Coupling (EMC) coefficient.

The first term in Equation (27) corresponds to polarization due to an applied voltage; the second is due to a pure piezoelectric effect; the third term is due to a pure strain gradient or flexoelectric effect (polarization exists even in the absence of applied voltage and piezoelectric effect as long as the strain is nonuniform) whereas the last two terms correspond to combined piezoelectric and flexoelectric contributions and thus informs us of the nonlinear interaction between flexoelectricity and piezoelectricity.

Note that our solution coincides with the results of Tadmor *et al.*[40] if we neglect the higher order contribution of polarization and strain gradients ($e \to 0$ and $f, f' \to 0$). In addition, if we further disregard the EMC ($\xi \to 0$), we recover the classical result for a simple dielectric in which the electric field is a constant $-\dfrac{V}{h}$.

To proceed further it is expedient to define the strain energy U:

$$U = \frac{1}{2} \iiint\limits_V T_{ij} S_{ij} dV + \frac{1}{2} \iiint\limits_V T_{ijm} u_{i,jm} dV \tag{28}$$

For the case of the 1-D beam it reduces to

$$U = -\frac{1}{2} \int\limits_{x=0}^{L} \hat{M}(x) \frac{d^2 w(x)}{dx^2} dx - \frac{1}{2} \int\limits_{x=0}^{L} \hat{P}(x) \frac{d^2 w(x)}{dx^2} dx \tag{29}$$

where

$$\hat{M}(x) = \iint\limits_A z\, T_1(x,z) dy dz \quad \text{and} \quad \hat{P}(x) = \iint\limits_A f\, P_3(x,z) dy dz \tag{30}$$

are the resultant moment and the higher order resultant moment respectively.

In the absence of body forces, the work done by external forces due only to transverse loading $q(x)$ is

$$W(x) = \int\limits_{x=0}^{L} q(x) w(x) dx \tag{31}$$

The total potential energy $\prod$ is obtained from Equations (29) and (31) as

$$\prod = U - W = -\frac{1}{2} \int\limits_{x=0}^{L} (\hat{M}(x) + \hat{P}(x)) \frac{d^2 w(x)}{dx^2} dx - \int\limits_{x=0}^{L} q(x) w(x) dx \tag{32}$$

Its first variation is derived in a similar form as given by reference[42]

$$\delta \prod = [-(\hat{M}(x) + \hat{P}(x)) \delta w'(x)]_0^L + [(\frac{d\hat{M}(x)}{dx} + \frac{d\hat{P}(x)}{dx}) \delta w(x)]_0^L$$

$$-\int\limits_0^L (\frac{d^2 \hat{M}(x)}{dx^2} + \frac{d^2 \hat{P}(x)}{dx^2} + q(x)) \delta w(x) dx \tag{33}$$

By use of the principle of minimum potential energy ($\delta \prod = 0$, e.g. reference[43]) and the fundamental lemma of calculus of variation (e.g. reference[44]) we have the following governing equation from Equation (33):

$$\frac{d^2 \hat{M}(x)}{dx^2} + \frac{d^2 \hat{P}(x)}{dx^2} + q(x) = 0, \qquad \forall\, x \in (0,L) \tag{34}$$

The corresponding boundary conditions prescribed at the beam ends (x=0 and x=*L*) are:

$$\begin{cases} \hat{M}(x) + \hat{P}(x) & \text{or} \quad \dfrac{dw(x)}{dx} \\[2ex] \dfrac{d(\hat{M}(x) + \hat{P}(x))}{dx} & \text{or} \quad w(x) \end{cases} \tag{35}$$

From Equations (30), (13) and (28) we can show that

$$\hat{M}(x) = -YI(1+\xi)\frac{d^2 w(x)}{dx^2}$$

$$\hat{P}(x) = -Af\left[\frac{V\varepsilon_0\chi}{h} + \frac{f'}{R(x)} + \frac{\xi^2(e-f)}{d^2 Y\, R(x)} + \frac{24V\varepsilon_0\chi\xi(e-f)}{dYh^2}\right] \tag{36}$$

where $I = \displaystyle\iint_A z^2 dA$ is the second moment of cross-sectional area A.

Thus, the equilibrium Equation (34) becomes

$$G\frac{d^4 w(x)}{dx^4} = q(x) \tag{37}$$

where G is the beam bending rigidity defined as

$$G = YI\left[1 + \xi + \frac{Aff'}{YI} + \frac{Af\xi^2(e-f)}{d^2 Y^2 I}\right] \tag{38}$$

Once again, we point out that if we ignore the polarization and strain gradients effects ($e \to 0$ and $f, f' \to 0$), we recover the same bending rigidity as in refrence[40]. Also, if we neglect the EMC ($\xi \to 0$), we recover the classical bending rigidity for a beam $G = YI$. Note that in the absence of piezoelectricity ($\xi \to 0$), the renormalized bending rigidity is $G = YI + Aff'$ due to flexoelectric effect.

The preceding derivation is for an arbitrary cross-sectional beam. As a concrete example, consider a rectangular cantilever beam subjected to a transversal point load N. The corresponding boundary conditions from Equation (35) are

$$w(0)=0 \quad \text{and} \quad \frac{dw(x)}{dx}\Big|_{x=0} = 0$$

$$\hat{M}(L) + \hat{P}(L) = 0 \quad \text{and} \quad \frac{d(\hat{M}(x) + \hat{P}(x))}{dx}\Big|_{x=L} = \text{N} \tag{39}$$

In the absence of distributed transverse loading ($q(x) = 0$) the homogeneous equilibrium equation becomes

$$G\frac{d^4 w(x)}{dx^4} = 0 \tag{40}$$

where the solution is in the form

$$G\, w(x) = \frac{a_1}{6}x^3 + \frac{a_2}{2}x^2 + a_3 x + a_4 \tag{41}$$

By means of the BC in Equation (39), the beam deflection is then,

$$w(x) = \frac{N\mathrm{x}^2(3L-x)}{6G} \tag{42}$$

in which the bending rigidity G is defined by Equation (38) with $I = \frac{bh^3}{12}$ and $A = bh$.

Thus, we may use the classical well-known beam equation for deflection provided the rigidity (or in effect the elastic modulus) is renormalized according to Equation (38).

## IV. PHYSICAL INSIGHTS, POSSIBILITY OF "GIANT" SIZE-DEPENDENT PIEZOELECTRICITY AND SCALING OF ELASTIC MODULUS

Based on the derivation in the preceding section, we may define an "effective" or "renormalized" piezoelectric constant which has contributions from both classical piezoelectricity and flexoelectricity.

From Equation (38), we define the effective coupling coefficient as:

$$\xi_{eff} = \xi + \frac{12f(e-f)\xi^2}{Y^2d^2h^2} + \frac{12ff'}{Yh^2} \tag{43}$$

Consequently, the "effective" piezoelectric coefficient is:

$$d_{eff} = \sqrt{\frac{\varepsilon}{Y}\frac{\xi_{eff}}{(1+\xi_{eff})}} \tag{44}$$

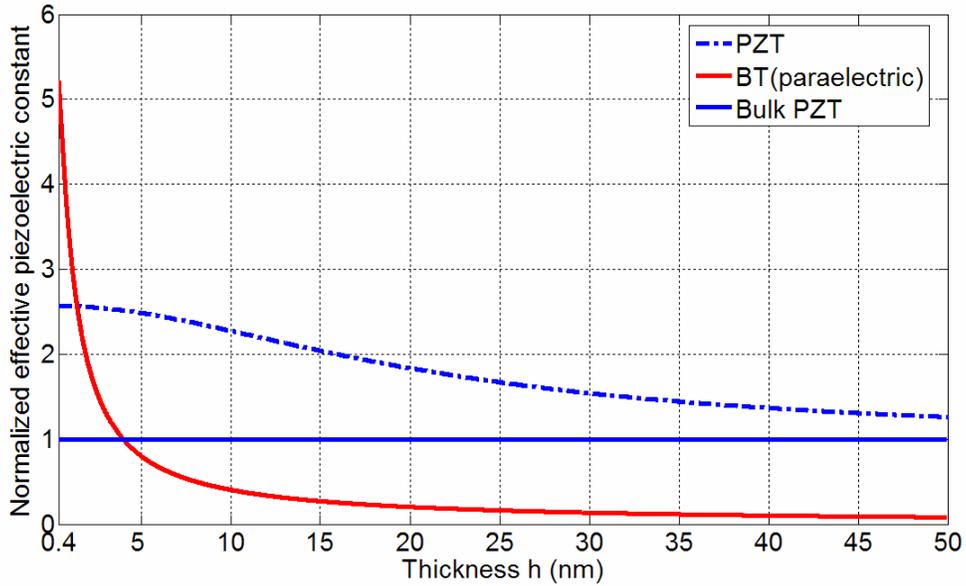

**Figure 3:** (Color online) Normalized effective piezoelectric constant of deformed PZT (dashed blue, dashed dark grey in print) and non-piezoelectric BT beams (solid red, light grey in print). The normalization is done with respect to the bulk piezoelectric constant of PZT (solid blue, dark grey in print) ( $d_{PZT} = -274 \ pC / N$ ) and piezoelectric phase of BT ( $d_{BT} = -78 \ pC / N$ ).

Figure (3) shows that for piezoelectric PZT cantilevers (dashed line), the effective piezoelectric constant is increased by 75% of the PZT bulk value ($d_{PZT} = -274\,pC/N$) at 20 nm. Even though cubic BT is not piezoelectric (red solid line), we still see a large apparent piezoelectric response below 10 nm. At 8 nm, the apparent piezoelectric response of BT is 50% that of the bulk BT piezoelectric constant ($d_{BT} = -78\,pC/N$). At 2 nm, the apparent piezoelectric response is double the one generated by a piezoelectric BT beam. An extremely high apparent piezoelectric response is seen at smaller sizes, reaching almost 5 times the piezoelectric BT constant.

Cross and co-workers report that ferroelectric phase (piezoelectric) BT has a high flexoelectric constant estimated[45] to be $f_{BT} = 50\,\mu C/m$. Figure (4) shows that for piezoelectric BT, the effective piezoelectric response increases by 20% of its bulk value at 8 $\mu m$ and exhibits a "giant" 500 % increase at 5 nm!

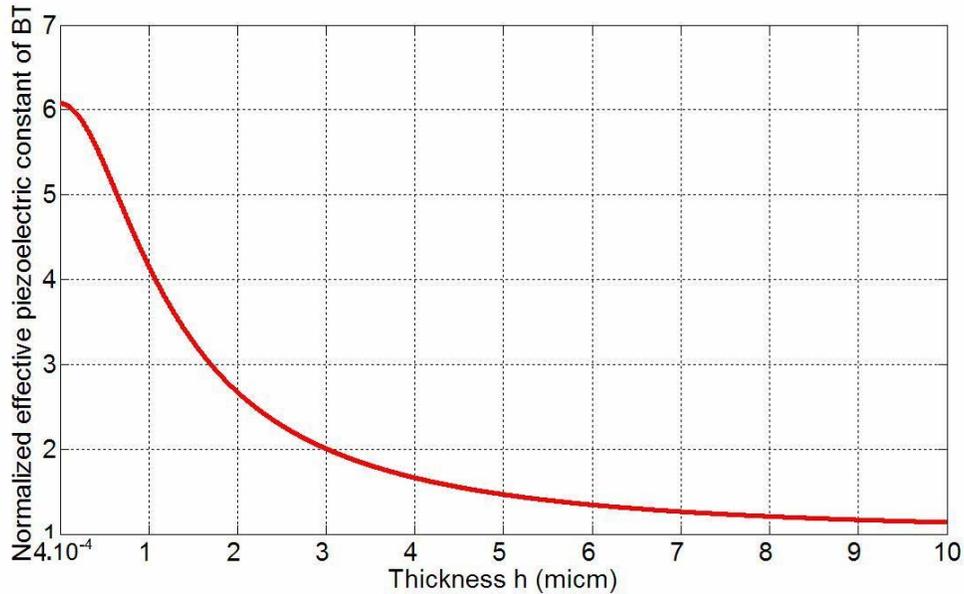

**Figure 4:** Normalized effective piezoelectric constant of tetragonal (piezoelectric) BT beam. An enhancement of 20% of its bulk value at 8 $\mu m$ and a 500 % increase at 5 nm is observed.

Our theoretical results indicate that the apparent piezoelectric response is determined by a synergistic addition between piezoelectricity and flexoelectricity (e.g. Equation 27). By comparing PZT and piezoelectric BT results, the noteworthy increase in the piezoelectric response occurs at vastly different length scales. The effective piezoelectric constant as defined previously in Equation (44) depends on both piezoelectric constant (EMC) and the flexoelectric constant. The piezoelectric constant for PZT is higher than BT but it is of the same order of magnitude. However, the flexoelectric constant of BT is two orders of magnitude higher than that of PZT ($f_{BT} = 100 f_{PZT}$) which explains the difference in the length scales at which the enhancement is observed. In the case of non-piezoelectric (paraelectric) BT, the only contribution to the effective piezoelectric response is due to flexoelectricity. Therefore, the effect is smaller and only occurs at small scales (10s nanometers).

We now show that flexoelectricity also impacts the observed or "apparent" elastic modulus. The normalized effective Young's modulus (with bulk value) is defined as (see Equation 38 also):

$$Y' = \frac{G}{YI} = [1 + \xi + \frac{12f(e-f)\xi^2}{Y^2d^2h^2} + \frac{12ff'}{Yh^2}] \qquad (45)$$

To illustrate our results, we pick the following values for the different parameters: A PZT 5H beam with rectangular cross-section defined by b=2h (b<5h, plane stress) and L=20h loaded with a force magnitude N=100 $\mu N$ so that we remain in the elastic domain. The flexoelectric coefficient $f$ is obtained from reference[12] $f_{PZT} = 0.5 \mu C/m$. A $1/h^2$ scaling is evident in Equation (45) and is illustrated in Figure (5): smaller beams appear stiffer due to the flexoelectric effect (dashed line). Note that the normalized effective Young's modulus according to Tadmor *et al.*[40] is little larger than one because of the EMC contribution.

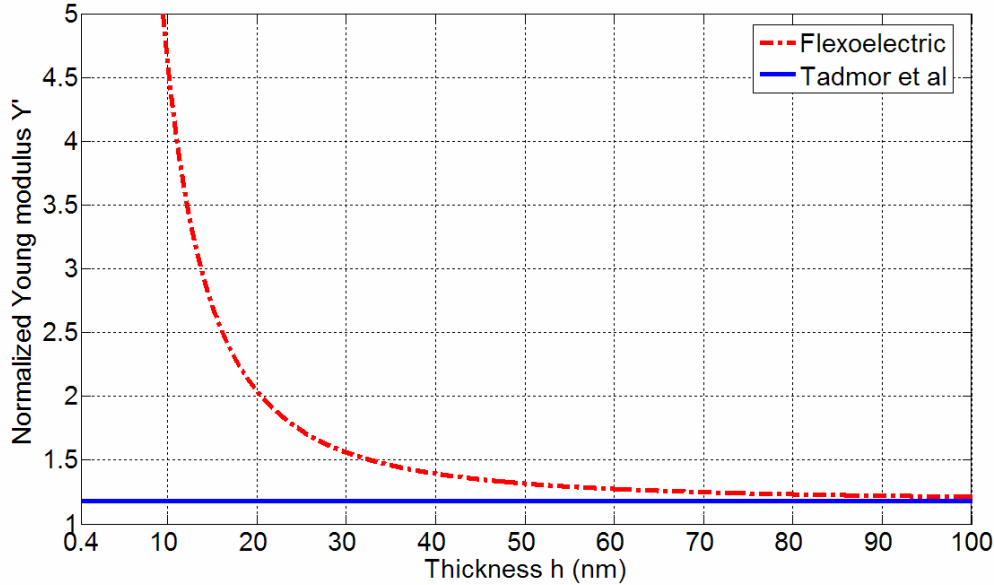

**Figure 5:** Normalized Young's modulus of a rectangular PZT cantilever beam. The dashed line illustrates the size dependency of the elastic modulus and exhibits a $1/h^2$ scaling where h is the beam thickness. The horizontal solid line is for the results of Tadmor *et al.* for classical piezoelectric beam that excludes the flexoelectric effect.

## V. ATOMISTIC SIMULATIONS

In previous sections, based on the phenomenon of flexoelectrcity and an appropriate mathematical description, we have argued the possibility of giant piezoelectricity is piezoelectric nanostructures and certainly an enhancement even in non-piezoelectric ones. In this section we present discrete atomistic calculations based on a (quantum mechanically derived) force field to confirm some of our predictions. We have avoided atomistic calculation of PZT since the core-shell potential available for it is not parameterized appropriately for the physical insights sought by the present work. Therefore we focused our attention mainly on BT. At temperature above the Curie temperature $T_C$ of 393K, BT is in its stable paraelectric cubic phase (Pm3m). Below $T_c$, BT undergoes three ferroelectric phase transitions. The cubic structure changes to tetragonal (P4mm) symmetry at $T_C$, orthorhombic (Amm2) at 278K and the last phase transition, rhombohedral (R3m) occurs at 183K. In prior work, one of us, Cagin *et al.*[46] has developed a suitable polarizable charge distribution Force Field for BT to use in molecular dynamics (MD) simulations based on *ab initio* quantum mechanical calculations. One distinctive feature of this force field is that charge transfer and atomic polarization are treated self-consistently and is thus quite appropriate for studying ferroelectrics. The charge is described as a Gaussian distribution for each of the core and the shell. The total core charge has positive fixed amplitude centered on the nucleus whereas the negative valence (shell) charge is determined via charge equilibration and is allowed to move off the nuclear center. The two Gaussian charge distributions interact with Coulombic (electrostatic) forces. Nonbonded interactions between neutral atoms and molecules (short range Pauli repulsion and long range attractive van der Waals dispersion) are described by the Morse potential. Our previous MD calculations[46] indicate that the polarizable charge distribution FF potential for BT is able to correctly predict experimentally observed paraelectric (cubic) to ferroelectric (tetragonal) phase transition among other features. One of us has also recently successfully used it to study antiferroelectricity in cubic and ferroelectric phases of BT[47].

Our calculations were carried out using MST package[48]. Beam thicknesses were varied from 1 single unit cell (4.01A as lattice parameter) to 2 nm while length was set to 4 nm. Several simulations were performed and results were averaged over all the runs. To reproduce our theoretical work conditions, simulated rectangular cantilevers were held fixed at one side then bent to the shape dictated by the simple 1D deflection solution defined previously in Equation (40) (Figure 6---deflection amplified to be seen).

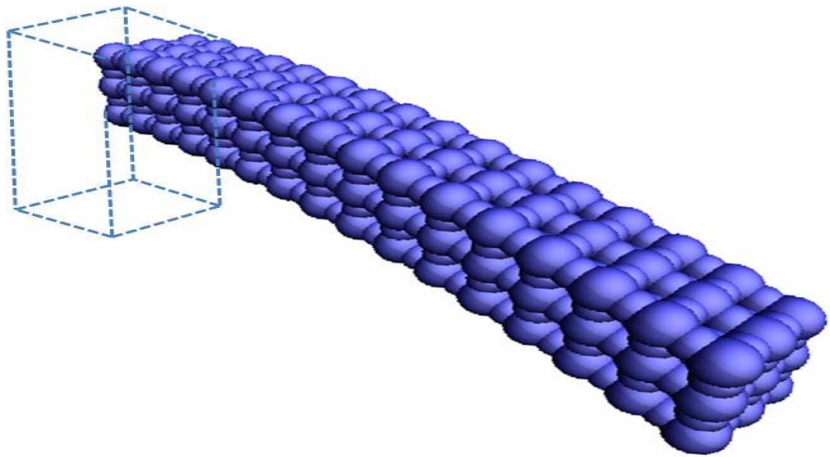

**Figure 6:** Atomistic representation of a cantilever beam under bending. The square dotted block is for aesthetic perspective.

For a given strain gradient, we determine the average polarization for different runs with different BT beam sizes in both ferroelectric and paraelectric phases. In the case of non-piezoelectric BT (Figure 7), MD calculations are in a good agreement with the predictions of our theoretical model from previous section. Only few points are calculated by MD are shown and the solid line is interpolated using the least square technique providing a guide to the eye.

For piezoelectric BT, as shown in the previous section, the effective piezoelectric response shows an enhancement at a higher length scale of few micrometers and reaches gigantic proportions at the nanoscale (Figure 8). Such "giant" enhancements are duly confirmed by atomistic calculations.

There are of course some (inconsequential) differences between the theoretical model and MD results. The theory is developed under the assumptions of simplified 1D problem whereas the simulations are carried out on 3D nanostructures. In addition, our model sensitively depends on several material properties (Young's modulus, dielectric, piezoelectric and flexoelectric constants) values of which could be over or under estimated by the experimental values we have used. At such small scales, other phenomena, in particular surface piezoelectricity/flexoelectricity[9], which are not taken into account by our model, may become important.

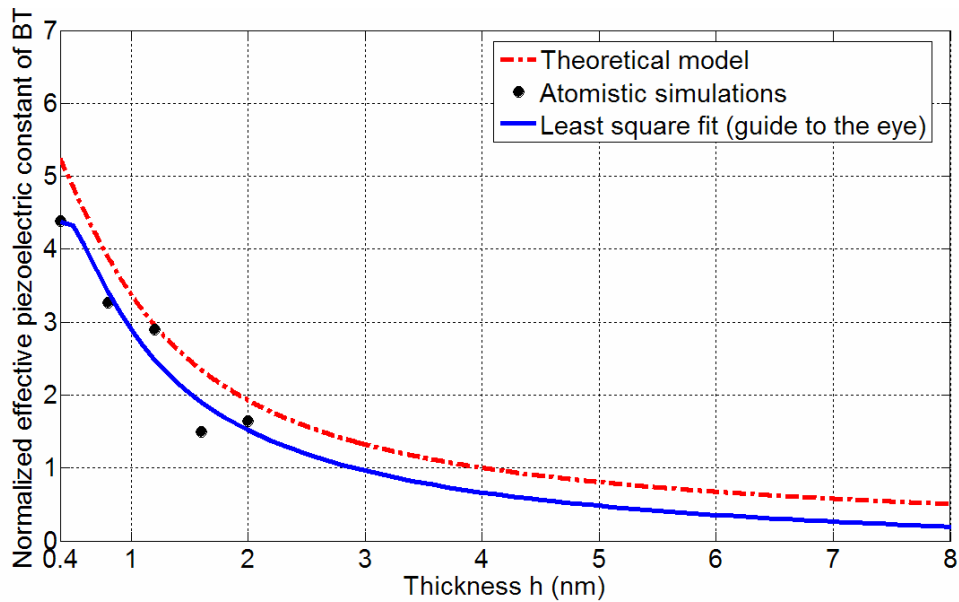

**Figure 7:** Normalized effective piezoelectric constant of cubic (non-piezoelectric) BT. Only a few points are obtained from atomistic simulations. The least square fit shows good agreement with the predictions of the theoretical model.

We have also computed and contrasted the effective elastic modulus with our theoretical results. The energy difference between the beam bent configuration and the undeformed one is the strain energy or the work done by the applied force. The strain energy U is:

$$U = \frac{1}{2}\int_V \sigma S\, dv = \frac{1}{2} YI \int_0^L \frac{1}{R^2(x)}\, dx \qquad (46)$$

We estimate the normalized effective Young's modulus from the following relation:

$$Y' = \sqrt{\frac{N^2 L^3}{6\, YI\ U}} \qquad (47)$$

We note that a similar technique was used by Miller and Shenoy[49] to explain atomistically the size dependency of the Young's modulus of nanosized elements and the flexural rigidity of beams in bending due to surface energy effects.

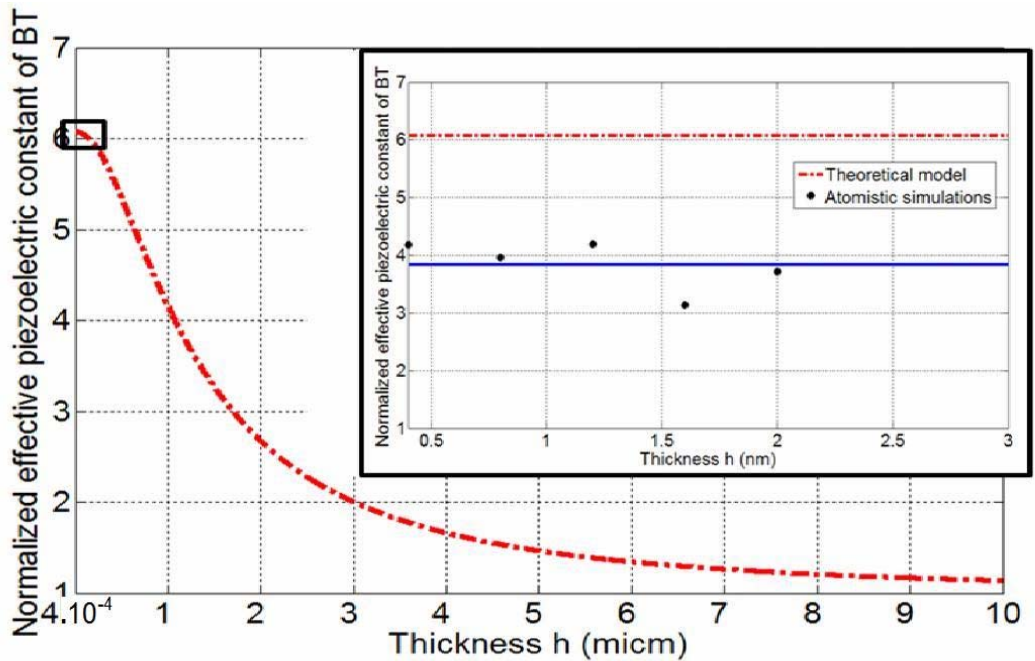

**Figure 8:** Normalized effective piezoelectric constant of tetragonal (piezoelectric) BT. Since the atomistic calculations were carried for very small sizes, the right inset corresponds to a zoomed-in view around 3 nm. The atomistic results fluctuate around a constant value (Least square fit (solid line)) and qualitatively match the theoretical predictions.

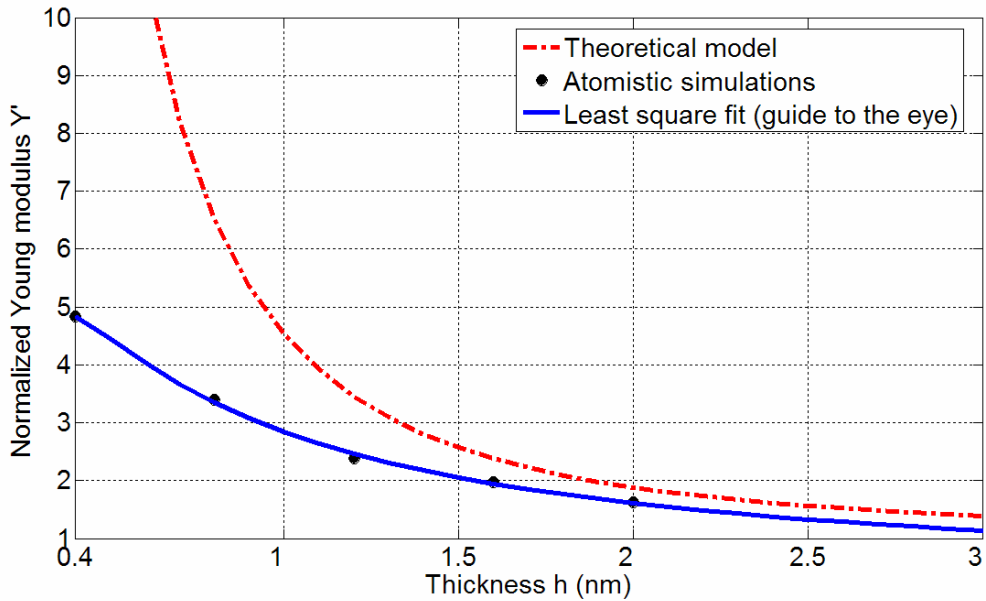

**Figure 9:** Normalized Young modulus for a BT cantilever beam. The least square fit of the atomistic simulations demonstrates reasonable agreement down to 2 nm.

The atomistic results for the Young's modulus for BT are contrasted wih theoretical ones in Figure (9). Once again, down to about 2 nm or so, there is is good agreement (and below which, as already explained, results diverge).

## VI. SUMMARY

We have argued that flexoelectricity exhibits a size-effect and thus should have important ramifications for the apparent piezoelectric and elastic behavior of nanostructures. Certainly in some dielectrics, flexoelectric coefficients are quite high and coupled with large strain gradients possible at the nanoscale, the effect of flexoelectricity can be non-trivial. In particular, using a model system of a cantilever nano-beam, we are able to analytically show that in materials that are already piezoelectric, the effect of flexoelectricity is multiplicative and combines nonlinearly with the intrinsic piezoelectricity. The nonlinear flexoelectric-piezoelectric interaction manifests itself as a "giant" increase in the apparent piezoelectric response at small sizes for materials that are intrinsically piezoelectric (duly confirmed via accurate atomistic calculations for BT). As is well-known in the classical piezoelectricity literature, a polarized elastic solid shows a renormalized (size-*independent*) elastic constant. This is true in flexoelectricity induced elasticity renormalization as well although the behavior is size-*dependent* and scales as ~$1/h^2$.

We find it interesting that classical piezoelectric theory when supplemented with flexoelectricity is able to capture the electromechanical behavior of nanostructures almost down to 2 nm's. Needless to say, without incorporation of flexoelectricity, the size-effects observed in the atomistic calculations cannot be reconciled. An auxiliary benefit of the present work, thus, is that continuum piezoelectricity duly supplemented with flexoelectricity may be employed to study nanoscale piezoelectricity in a computationally expedient manner rather than using atomistic calculations which have clear computational limits in terms of system size and computational expense.

Currently very little experimental work is available on piezoelectricity bent nano-beam as it is highly challenging to perform controlled experiments at that scale. In that regard we note that in some cases (e.g. in piezoelectric phase of $BaTiO_3$) the size effect predicted by us are also manifest at micron size beams thus providing a facile route for experimental verification of our presented scaling laws. Furthermore, the approach and conclusions of this work will remain relevant for same order of magnitude structures such as lattice-mismatched epitaxial thin films[50]. The latter work examined the influence of strain gradients (through flexoelectric coupling) on the ferroelectric properties of films with decreasing thickness. Another example is the case of asymmetric three-component ferroelectric superlattices[51] where authors confirm enhancement in polarization by similar phenomena (breaking the inversion symmetry of the lattice).

Our theoretical model neglects some effects that may become important at small sizes e.g. surface flexoelectricity and surface piezoelectricity[9]. Regarding the latter, we have minimized its influence in atomistic calculations by ensuring the centro-symmetry of surfaces. Surface flexoelectricity has been discussed at length by Tagantsev[9] and is not included in our theoretical model (although this phenomenon is automatically accounted for in the atomistic calculations). Evidently, surface flexoelectricity is likely to be important only below 2 nms or so for the materials we have investigated (given the close agreement up to that point between our atomistic and theoretical results).

## Acknowledgments

Majdoub acknowledges useful discussions with Takahiro Shimada on PZT core shell potential. Authors are also grateful to Dr Gillian C. Lynch for use of Cerius2 software to handle some aspects of the calculations. Sharma and Majdoub acknowledge financial support from NSF NIRT grant CMMI 0708096 (program manager Clark Cooper) and Texas ARP.